\documentclass[twocolumn,trackchanges]{aastex631}

\usepackage{graphicx}
\usepackage{multirow}
\usepackage{amsmath}
\usepackage{newtxtext}
\usepackage{newtxmath}
\usepackage{textalpha}
\usepackage[style=english]{csquotes}

\usepackage{siunitx}
\newcommand{\circa}{\ensuremath{{\sim}\mspace{2mu}}}
\newcommand{\ut}{\textsc{ut}}
\DeclareSIPostPower{\nominal}{N}
\DeclareSIQualifier{\earth}{\ensuremath{\oplus}}
\DeclareSIQualifier{\jupiter}{J}
\DeclareSIQualifier{\planet}{p}
\DeclareSIQualifier{\etoile}{\ensuremath{\star}}
\DeclareSIQualifier{\sun}{\ensuremath{\odot}}
\DeclareSIUnit{\density}{\ensuremath{\mathnormal{\rho}}}
\DeclareSIUnit{\erg}{erg}
\DeclareSIUnit{\luminosity}{\ensuremath{\mathnormal{L}}}
\DeclareSIUnit{\magnitude}{mag}
\DeclareSIUnit{\mass}{\ensuremath{\mathnormal{M}}}
\DeclareSIUnit{\mas}{\milliarcsecond}
\DeclareSIUnit{\milliarcsecond}{mas}
\DeclareSIUnit{\parsec}{pc}
\DeclareSIUnit{\radius}{\ensuremath{\mathnormal{R}}}
\DeclareSIUnit{\year}{yr}

\newcommand{\pmra}{\ensuremath{\mu_\alpha \cos\delta}}
\newcommand{\pmdec}{\ensuremath{\mu_\delta}}

\newcommand{\tempeff}{\ensuremath{T_{\mathrm{eff}}}}

\newcommand{\uniformdist}{\mathcal{U}}
\newcommand{\normaldist}{\mathcal{N}}

\newcommand{\kepler}{Kepler}
\newcommand{\ktwo}{K2}
\newcommand{\tess}{TESS}
\newcommand{\gaia}{Gaia}

\newcommand{\gaiaG}{\ensuremath{G}}
\newcommand{\gaiaBP}{\ensuremath{G_\mathrm{BP}}}
\newcommand{\gaiaRP}{\ensuremath{G_\mathrm{RP}}}

\newcommand{\BPminusRP}{\ensuremath{\gaiaBP - \gaiaRP}}

\newcommand{\celeritetwo}{\textsc{celerite2}}

\newcommand{\healpix}{HEALPix}

\newcommand{\pymc}{\textsc{PyMC}}

\newcommand{\planet}[2]{#1\thinspace #2}
\newcommand{\blancoone}{Blanco\thinspace 1}

\newcommand{\contamratio}{0.33}

\received{April 18, 2024}
\revised{August 29, 2024}
\accepted{September 5, 2024}
\submitjournal{\apj}

\shorttitle{Confirming the Tidal Tails of Blanco 1}
\shortauthors{Sha et al.}

\graphicspath{{./}{plot/}}

\begin{document}

\title{%
Confirming the Tidal Tails of the Young Open Cluster Blanco 1
with TESS Rotation Periods
}

\correspondingauthor{Lizhou Sha}
\email{lsha@wisc.edu}

\author[0000-0001-5401-8079]{Lizhou Sha}
\affiliation{Department of Astronomy, University of Wisconsin--Madison, 475 N Charter St, Madison, WI 53706 USA}

\author[0000-0001-7246-5438]{Andrew M. Vanderburg}
\affiliation{Department of Physics and Kavli Institute for Astrophysics and Space Research, Massachusetts Institute of Technology, Cambridge, MA 02139, USA}

\author[0000-0002-0514-5538]{Luke G. Bouma}
\affiliation{Department of Astronomy, California Institute of Technology, Pasadena, CA 91125, USA}

\author[0000-0003-0918-7484]{Chelsea X. Huang}
\affiliation{Centre for Astrophysics, University of Southern Queensland, Toowoomba, Queensland 4350, Australia }

\begin{abstract}
\blancoone{} is an $\approx \SI{130}{\mega\year}$ open cluster located
\SI{240}{\parsec} from the Sun
below the Galactic plane.
Recent studies have reported the existence of diffuse tidal tails extending
\SIrange[range-phrase=--]{50}{60}{\parsec} from the cluster center
based on the positions and velocities measured by \gaia{}. 
To independently assess the reality and extent of this structure,
we used light curves generated from \tess{} full-frame images to search for photometric rotation periods of stars in and around \blancoone{}. 
We detected rotation periods
down to a stellar effective temperature of $\approx \SI{3100}{\kelvin}$
in 347 of the 603 cluster member candidates for which we have light curves.
For cluster members in the core and candidate members in the tidal tails,
both within a temperature range of \SIrange{4400}{6200}{\kelvin},
74\% and 72\% of the rotation periods
are consistent with the single-star gyrochronological sequence, respectively. 
In contrast, a comparison sample of field stars yielded gyrochrone-consistent rotation periods for only 8.5\% of stars.
The tidal tail candidates' overall conformance to the core members' gyrochrone sequence implies that their contamination ratio is consistent with zero
and $<\contamratio$ at the $2\sigma$ level.
This result confirms the existence of \blancoone{} tidal tails
and doubles the number of \blancoone{} members
for which there are both spatio-kinematic and rotation-based cluster membership verification.
Extending the strategy of using \tess{} light curves for gyrochronology to other nearby young open clusters and stellar associations
may provide a viable strategy for
mapping out their dissolution
and broadening the search for young exoplanets.
\end{abstract}


\section{Introduction} \label{sec:intro}

The abundance of precise astrometric, kinematic, and photometric data brought to us by the \gaia{} mission
\citep{2016A&A...595A...1G}
has ushered in a new age of knowledge in stellar clusters and associations.
With discoveries like
the Gaia--Enceladus sausage \citep{2018MNRAS.478..611B},
the Theia groups \citep{2019AJ....158..122K,2021ApJ...915L..29G},
and tidal tails or coronae of open clusters \citep{2018A&A...618A..93C,2021A&A...645A..84M},
\gaia{} has enriched our story of Galactic structure.
Given \gaia{} DR3's completeness down to a magnitude of $\gaiaG = 20.7$
\citep{2021A&A...652A..76H},
we can leverage its data to trace the dissolution of clusters
spanning a large portion of the sky.
This all-sky coverage is critical because
stars form bound and embedded in their primordial molecular clouds,
but as they turn on and disperse the clouds,
the cluster becomes \enquote{open}
and dissolves
over a typical timescale of \SIrange[range-phrase=--]{10}{100}{\mega\year}
\citep{2003ARA&A..41...57L,2019ARA&A..57..227K}.
Thus, by leveraging \gaia{}'s extreme precision and all-sky coverage,
we are gaining a much more complete picture of diffuse structures like
stellar associations and tidal tails of open clusters than previously possible.
While \gaia{}'s ability to detect diffuse structures is revolutionizing the understanding of our Galaxy,
studying these objects poses unique challenges.
In particular, because these diffuse structures span significantly larger volumes
in the Galaxy than compact structures like open clusters, 
the probability that field stars
can mimic the kinematic and color--magnitude signatures of association members increases greatly.
Thus, \gaia{} data alone are often not sufficient for validating membership.
This problem is exacerbated by different clustering algorithms, such as hierarchical density-based clustering (e.g. HDBSCAN) or Gaussian mixture models,
often yielding different extents for these associations \citep{2021A&A...646A.104H}.
As a result, we have to rely on additional age proxies,
such as stellar spin down due to magnetic braking,
lithium depletion boundary,
and coronal X-ray emission
to verify membership
(see \citealt{2014prpl.conf..219S} for a review).

Meanwhile, advances in high-cadence photometric surveys have made
using stellar rotation as a proxy for age, a practice known as gyrochronology
\citep{2003ApJ...586..464B},
an attractive observational strategy.
Gyrochronology studies of clusters and associations and exoplanet transit surveys
both benefit from uninterrupted monitoring of a wide area of the sky,
a strategy pioneered by the Monitor project
\citep{2007MNRAS.375...29A,2009MNRAS.392.1456I}.
After early successes from ground-based transit surveys such as HATNet \citep{2010MNRAS.408..475H} and SuperWASP \citep{2011MNRAS.413.2218D}
and the space-based \kepler{} mission \citep{2011ApJ...733L...9M},
this strategy matured by the \ktwo{} mission
and was used to study the Pleiades \citep{Rebull:2016},
Hyades \citep{Douglas:2016},
and M67 \citep{2016ApJ...823...16B,2018ApJ...859..167E} clusters.
As the successor to \kepler{},
the Transiting Exoplanet Survey Satellite \citep[\tess{};][]{2015JATIS...1a4003R}
trades off spatial resolution for a much wider field of view,
allowing us to efficiently apply gyrochronology
to larger and more diffuse structures.
To wit, \citet{2019AJ....158...77C} verified the existence of
the Pisces--Eridanus stream \citep{2019A&A...622L..13M} and found it to have the same age as the Pleiades.
Later studies have verified the tidal tails around NGC\thinspace 2516 \citep{2021AJ....162..197B}
and a diffuse complex of stars in the \textalpha\thinspace Persei cluster \citep{2023AJ....166...14B}.
Furthermore, the search for young planets in \tess{} and \kepler{} data
has also led to the discovery of new young stellar associations
\citep{2021AJ....161..171T,2022AJ....164...88B,2022AJ....164..115N}.

It is in light of the expanded cluster membership knowledge in the wake of \gaia{}
and the precise photometry over a large field of view by \tess{}
that we revisit rotation periods in the
open cluster \object[Blanco 1]{\blancoone{}}.
This
nearby ($\approx \SI{240}{\parsec}$, \citealt{2021ApJ...912..162P})
and
young ($\approx \SI{130}{\mega\year}$, \citealt{2010ApJ...725L.111C})
cluster has been extensively studied through photometry, spectroscopy, and astrometry
in the 75 years since its discovery by \citet{1949PASP...61..183B}.
The first ground-based transit survey to study \blancoone{} was
Kilodegree Extremely Little Telescope--South
\citep[KELT-South;][]{2008AJ....135..907P,2012PASP..124..230P},
which monitored 33 stars for 43 nights spread over 90 days and detected periods for 23 of them
\citep{2014ApJ...782...29C}.
Since \gaia{} DR2, the Next Generation Transit Survey (NGTS) had monitored 170 stars for $\approx 200$ nights and detected periods for 127
\citep{2020MNRAS.492.1008G}.
After those surveys had been published,
several authors reported on \blancoone{}'s extended structure,
variously known as \enquote{coronae} or \enquote{tidal tails}
\citep{2020ApJ...889...99Z,2021A&A...645A..84M,2021ApJ...912..162P}.
Leveraging these expanded member lists,
we aim to extend these earlier rotation period studies to stars in \blancoone{}'s tidal tails
and, for the first time, independently verify that their age is consistent
with stars in the cluster core.

This paper is organized as follows.
We begin by detailing our procedure for selecting candidate members of \blancoone{}
and a matched comparison sample of field stars for experimental control
in \autoref{sec:cluster}.
After briefly explaining how we generated the \tess{} light curves in \autoref{sec:cdips},
we describe our method for measuring rotation periods in \autoref{sec:period}.
These measurements result (\autoref{sec:results})
in a stellar effective temperature--rotation period relation for \blancoone{}
that contains the tidal tail candidates for the first time,
and we use this relation
to estimate the field star contamination rate in the tidal tails.
Finally, in \autoref{sec:discussion},
we round off this paper by discussing the implications of the verified tidal tails,
both for the morphology of \blancoone{}
and the possibility of extending this observational strategy to other nearby young open clusters,
focusing on the prospect for enhancing the local young star census
and the search for young exoplanets.

\section{Selecting Cluster Candidates and a Comparison Sample} \label{sec:cluster}

\begin{figure}
    {
    \centering
    \includegraphics[width=\columnwidth]{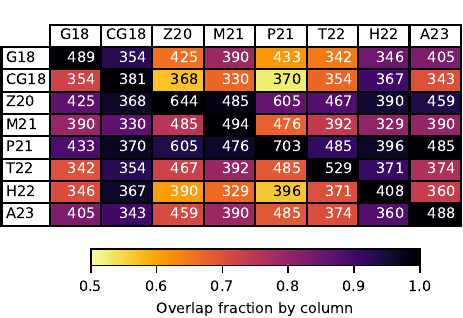}
    \caption{A table summarizing the degree to which eight literature references
    are in agreement over \blancoone{} candidates.
    Each row and column represent one of
    eight references that contain membership lists for \blancoone{} compiled using \gaia{}
    DR2 or (E)DR3 data.
    The number in each cell is the number of candidates shared by that row's and column's references.
    The color in each cell indicate the fraction of common candidates with respect to the total number identified by each column's reference.
    \label{fig:target_overlap}
    }
    }

    \textbf{References.}
    G18: \citet{2018A&A...616A..10G};
    CG18: \citet{2018A&A...618A..93C};
    Z20: \citet{2020ApJ...889...99Z};
    M21: \citet{2021A&A...645A..84M};
    P21: \citet{2021ApJ...912..162P};
    T22: \citet{2022A&A...659A..59T};
    H22: \citet{2022ApJS..262....7H};
    A23: \citet{2023A&A...677A.163A}.
\end{figure}

\begin{figure*}
    \centering
    \includegraphics[width=\textwidth]{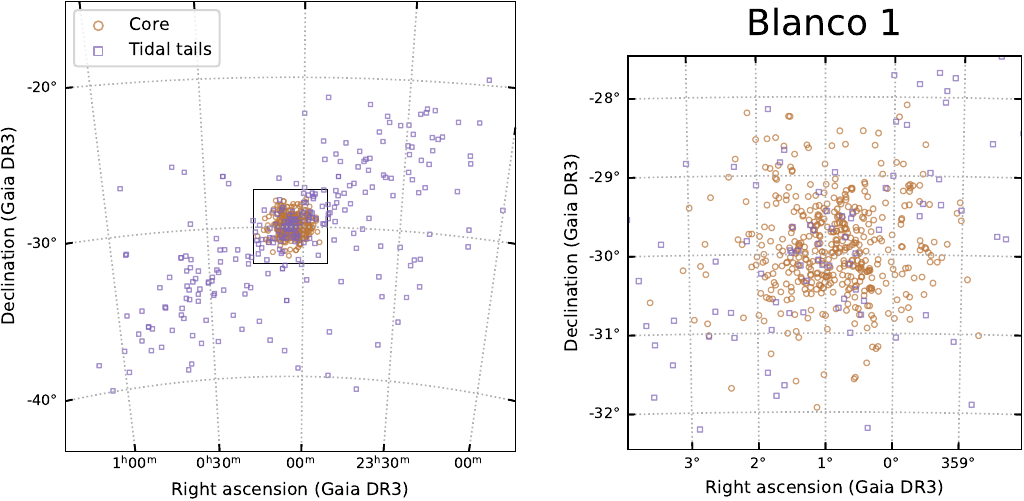}
    \caption{Sky charts for \blancoone{} cluster candidate members identified by \citet{2021ApJ...912..162P}.
    The charted positions are from \gaia{} DR3.
    \textbf{Left:} Full view of \blancoone{}.
    \textbf{Right:} Zoomed-in view of the cluster core.
    The circular orange marks are the stars within the tidal radius of the cluster,
    and the square magenta marks are those outside of the tidal radius.}
    \label{fig:skymap}
\end{figure*}

\subsection[Selecting a candidate list of Blanco 1 members]{Selecting a candidate list of \blancoone{} members}

In selecting a candidate list of \blancoone{} membership for this paper,
we focused on those produced in recent years based on \gaia{} DR2 or (E)DR3 data.
Together with the \gaia{} DR2 catalog
\citep{2018A&A...616A...1G},
\cite{2018A&A...616A..10G} released a list of 489 \blancoone{} candidates based on astrometry.
Using the unsupervised UPMASK algorithm, \citet{2018A&A...618A..93C} found 381 candidates
with membership probabilities over 50\%.
Still using \gaia{} DR2 but with a larger search radius of \SI{100}{\parsec},
\citet{2020ApJ...889...99Z} identified the tidal tail structure of \blancoone{}
for the first time with the unsupervised machine learning algorithm \textsc{StarGO},
listing 644 candidates.
\citet{2021A&A...645A..84M} confirmed the existence of the tidal tails
(which they called \enquote{coronae}) and listed 494 candidates,
taking into account cluster bulk velocities and correcting for line-of-sight distance errors.
Taking advantage of the higher astrometric precision of \gaia{} EDR3,
\citet{2021ApJ...912..162P} used \textsc{StarGO} to find 703 candidates,
the most thus far.
The most recent membership lists produced with \gaia{} (E)DR3
largely confirmed the earlier findings of the tidal tails,
with the number of cluster candidates ranging 408--529
\citep{2022A&A...659A..59T,2022ApJS..262....7H,2023A&A...677A.163A}.

We chose to adopt the membership list of \citet{2021ApJ...912..162P}
rather than attempting to synthesize a superset of all available literature references.
We found their list to be the most complete, encompassing 85\% of the 829 distinct candidates in the literature.
Further, almost 95\% of the 703 candidates identified by \citeauthor{2021ApJ...912..162P}
also appear in at least another reference,
whereas out of the 126 candidates not found on their list,
over 75\% only appear in one reference.
This high degree of overlap is visible in \autoref{fig:target_overlap}
as distinctly dark colors in the corresponding row labeled P21,
indicating that virtually all candidates found by \citeauthor{2021ApJ...912..162P}
are also identified by the authors corresponding to each column,
suggesting that their list has a low false positive rate.
For the sake of consistency and uniformity in selection criterion,
we adopt the 703 candidates of \citeauthor{2021ApJ...912..162P} for further analysis in the paper.
\autoref{fig:skymap} shows a chart of these candidates on the sky.

\begin{figure}
    \centering
    \includegraphics[width=\columnwidth]{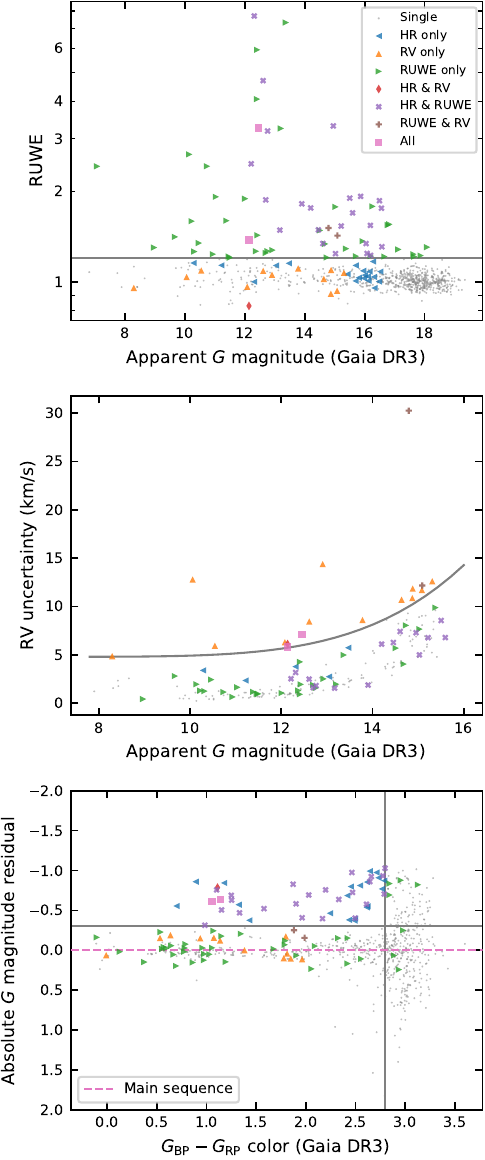}
    \caption{\gaia{} DR3 multiplicity indicators for \blancoone{} member candidates.  Stars that are apparently single appear in gray. The acronyms in the legend are as follows.
    HR: On the binary sequence in the Hertzsprung--Russell diagram.
    RV: Above the cutoff in radial velocity uncertainty.
    RUWE: Above the cutoff in \gaia{} DR3 Renormalised Unit Weight Error parameter.
    }
    \label{fig:binarity}
\end{figure}

\subsection{Stellar multiplicity}

As we would like to use stellar rotation periods to verify cluster membership,
we need to identify binary stars in the sample.
In the same cluster, binary stars often rotate faster than single stars
due to various mechanisms that inhibit angular momentum loss
\citep{2007ApJ...665L.155M}.
In addition, they may be eclipsing binaries or show ellipsoidal variations
that may be confused with rotational signal.
Here, we use the \gaia{} DR3 to systematically identify
astrometric, spectroscopic, and photometric binaries.
As detailed in the following paragraphs, we adopted three simple criteria for stellar multicity:
\begin{enumerate}
    \item Excess astrometric uncertainty: $\mathrm{RUWE} > 1.2$,
    \item Excess radial velocity (RV) uncertainty,
    \item Location in the binary sequence on the Hertzsprung--Russell diagram.
\end{enumerate}
Any candidate that satisfies any one of these criteria is deemed a likely multiple star for our analysis.

We identified astrometric binaries by the excess uncertainty in their \gaia{} DR3 astrometric solution.
This excess uncertainty is parameterized by the
unitless Renormalised Unit Weight Error (RUWE) parameter,
which is expected to be close to $1$ for \enquote{well-behaved} single sources
\citep{2021A&A...649A...2L}.
Based on the empirical distribution of RUWE of the \blancoone{} candidates (\autoref{fig:binarity}, top panel),
we chose a cutoff of
\begin{equation}
    \mathrm{RUWE} > 1.2
\end{equation}
for multiplicity.
Sources that satisfied this condition have the \texttt{multi\_ruwe\_flag} set to \texttt{True}
in Tables~\ref{tab:blanco1} and \ref{tab:control}.

We used \gaia{} DR3 RV uncertainties to identify spectroscopic binaries,
whenever \gaia{} DR3 RV measurements were available.
Based on the empirical dependence of RV uncertainty with photometric magnitude
(\autoref{fig:binarity}, middle panel),
we used a hand-tuned multiplicity cutoff in the form of
\begin{equation}
    \frac{\sigma_\mathrm{RV}}{\si{\kilo\meter\per\second}} >
    4.8 + 0.02 \, (G - 8)^2 + 0.002 \, (G - 8)^4 ,
\end{equation}
where $G$ is the \gaia{} \gaiaG{} magnitude of the star.
Sources that satisfied this condition have the \texttt{multi\_rv\_flag} set to \texttt{True}
in Tables~\ref{tab:blanco1} and \ref{tab:control}.

\begin{figure*}
    \centering
    \includegraphics[width=\textwidth]{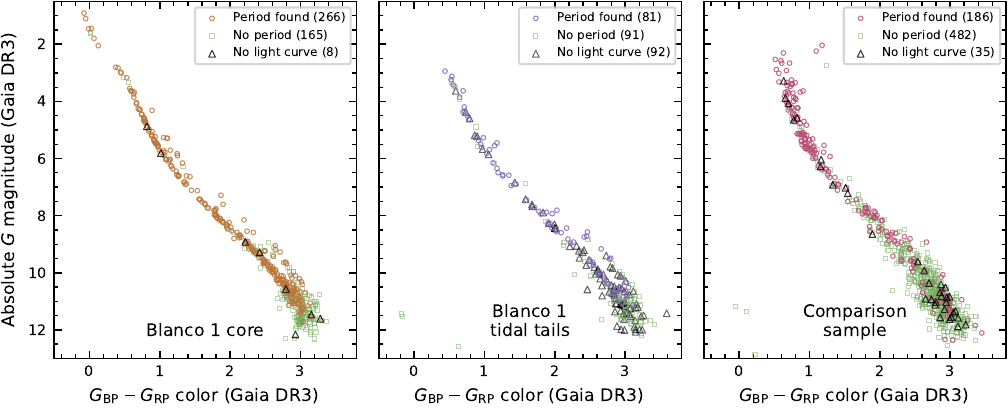}
    \caption{Hertzsprung--Russell diagrams of the open cluster \blancoone{}
        (\textbf{left}: core, \textbf{center}: tidal tails)
        and the comparison sample of field stars (\textbf{right}).
        Stars with measured rotation periods are plotted in circular marks of various colors,
        and those whose light curve show no identifiable rotation signal are plotted in square purple marks.
        The gray triangular marks are the stars without valid light curves in the sample.
    }
    \label{fig:hr}
\end{figure*}

Finally, we used the Hertzsprung--Russell (HR) diagram to identify \blancoone{} stars on the binary sequence.
The HR diagram (\autoref{fig:hr}) was constructed using \gaia{} DR3 \BPminusRP{} color and \gaiaG{} magnitudes,
which were converted to absolute magnitudes using the corrected distances from \citet{2021ApJ...912..162P}.
We fitted a 9th-order polynomial to the main sequence of \blancoone{} (see \autoref{sec:ms_poly} for details).
We deemed any stars with $\BPminusRP{} < 2.8$
and more than 0.3 magnitudes brighter than the main sequence as predicted by the 9th-order polynomial
to be part of the binary sequence
(\autoref{fig:binarity}, bottom panel).
The additional constraint on color was to exclude
M dwarfs that might not have converged on the zero-age main sequence at the age of \blancoone{}.
\blancoone{} stars deemed to be on the binary sequence had their \texttt{multi\_hr\_flag} set to \texttt{True}
in \autoref{tab:blanco1}.
As the comparison sample stars were not expected to be on the same isochrone,
we did not check them for the possibility of photometric binaries.

Altogether,
100 sources in the \blancoone{} candidate list from \citet{2021ApJ...912..162P} were marked as binaries.
Out of those, 67, 17, and 46 sources were marked according to at least one of the RUWE, RV, and HR criteria, respectively.
Furthermore, 4 sources were marked for at least RUWE and RV, 3 sources were marked for at least RV and HR,
25 sources were marked for at least RUWE and HR, and 2 sources were marked for all three criteria.

\subsection{Estimating effective temperatures}
\label{sec:teff}

We follow \citet{2020ApJ...904..140C} in estimating the effective temperature \tempeff{}
of \blancoone{}
by adopting a polynomial that approximates conversion between
dereddened \gaia{} DR2
\BPminusRP{} color \citep{2018A&A...616A...4E}
and \tempeff{}.
This empirical color--\tempeff{} relation is derived from benchmark stars reported by
\citet{2016ApJS..225...32B}, \citet{2012ApJ...757..112B}, and \citet{2015ApJ...804...64M}
and the scatter implies a \tempeff{} precision $\approx \SI{50}{\kelvin}$.
To deredden the observed \gaia{} DR2 \BPminusRP{} colors of our targets,
we adopt $E(B - V) = 0.010$ from \citet{2018A&A...616A..10G} as the color excess for \blancoone{},
which 
we translate into \gaia{} DR2 $E(\BPminusRP{})$ color excess
using the standard conversion $A_V = 3.1 E(B - V)$
and the empirical relation $E(\BPminusRP) = 0.415 A_V$
from \citet{2020ApJ...904..140C}.
In carrying out these calculations, we used a Python implementation by \citet{2023ApJ...947L...3B}.
The resulting temperatures are listed in \autoref{tab:blanco1} for \blancoone{} candidates 
and are used to derive the temperature--rotation period relationship in \autoref{sec:color_period}.

{\catcode`\&=11
\gdef\citelindengrengaiadrthree{\citet{2021A&A...649A...2L}}
\gdef\citegaiatwo{\citet{2018A&A...616A...1G}}
}
\begin{deluxetable*}{lccl}
    \tablecaption{Columns for the Table of \blancoone{} members
    \label{tab:blanco1}}
    \tablehead{%
        \multicolumn{1}{l}{Column} &
        \colhead{Unit} &
        \colhead{Data type} &
        \multicolumn{1}{l}{Description}
    }
    \startdata
    dr3\_source\_id & \ldots{} & int64 & \gaia{} DR3 source ID \\
    dr2\_source\_id & \ldots{} & int64 & \gaia{} DR2 source ID \\
    tic\_id & \ldots{} & int64 & \tess{} Input Catalog (TIC) ID (TIC 8.2) \\
    ra & deg & float64 & \gaia{} DR3 right ascension (J2016.0) \\
    ra\_error & deg & float64 & \gaia{} DR3 right ascension uncertainty (J2016.0) \\
    dec & deg & float64 & \gaia{} DR3 declination (J2016.0) \\
    dec\_error & deg & float64 & \gaia{} DR3 declination uncertainty (J2016.0) \\
    pmra & \si{\mas\per\year} & float64 & \gaia{} DR3 proper motion in right ascension (J2016.0) \\
    pmra\_error & \si{\mas\per\year} & float64 & \gaia{} DR3 proper motion uncertainty in right ascension (J2016.0) \\
    pmdec & \si{\mas\per\year} & float64 & \gaia{} DR3 proper motion in declination (J2016.0) \\
    pmdec\_error & \si{\mas\per\year} & float64 & \gaia{} DR3 proper motion uncertainty in declination (J2016.0) \\
    phot\_g\_mean\_mag & mag & float32 & \gaia{} DR3 \gaiaG{} magnitude \\
    bp\_rp & mag & float32 & \gaia{} DR3 \BPminusRP{} color \\
    phot\_g\_abs\_mag & mag & float32 & \gaia{} DR3 absolute \gaiaG{} magnitude, using distances from P21 \\
    ruwe & \ldots{} & float32 & \gaia{} DR3 Renormalised Unit Weight Error \\
    radial\_veolicty & \si{\km\per\s} & float32 & \gaia{} DR3 radial velocity \\
    radial\_veolicty\_error & \si{\km\per\s} & float32 & \gaia{} DR3 radial velocity uncertainty \\
    teff & K & float32 & Stellar effective temperature derived via the relation of \citet{2020ApJ...904..140C} (\autoref{sec:teff}) \\
    in\_rt & \ldots & bool & Flag for whether the target is within the tidal radius,
        using Galactic $XYZ$ coordinates from P21 \\
    multi\_ruwe\_flag & \ldots{} & bool & Flag for astrometric multiplicity as determined by \gaia{} DR3 RUWE \\
    multi\_rv\_flag & \ldots{} & bool & Flag for spectroscopic multiplicity as determined by \gaia{} DR3 RV uncertainty \\
    multi\_hr\_flag & \ldots{} & bool & Flag for photometric multiplicity as determined by \gaia{} DR3 color--magnitude diagram \\
    has\_cdips & \ldots & bool & Whether the target has a valid CDIPS light curve \\
    ls\_period & d & float64 & Rotation period determined by Lomb--Scargle periodogram \\
    ls\_flags & \ldots{} & string & Flags used in the vetting process (\autoref{sec:ls}) for Lomb--Scargle periods \\
    gp\_period & d & float64 & Median rotation period of the Gaussian process (GP) model posterior \\
    gp\_perioderr1 & d & float64 & Upper uncertainty on the rotation period of the GP model posterior \\
    gp\_perioderr2 & d & float64 & Lower uncertainty on the rotation period of the GP model posterior \\
    \enddata
    \tablecomments{The full table is available online in machine-readable format.
    The column descriptions are included here to demonstrate the table's content.}
    \tablereferences{\gaia{} DR3: \citelindengrengaiadrthree.
    \gaia{} DR2: \citegaiatwo.
    TIC~8.2: \citet{tic8,tic82}.
    P21: \citet{2021ApJ...912..162P}.
    }
\end{deluxetable*}

\begin{deluxetable*}{lccl}
    \tablecaption{Columns for the Table of Comparison Sample of Field Stars
    \label{tab:control}}
    \tablehead{%
        \multicolumn{1}{l}{Column} &
        \colhead{Unit} &
        \colhead{Data type} &
        \multicolumn{1}{l}{Description}
    }
    \startdata
    dr3\_source\_id & \ldots{} & int64 & \gaia{} DR3 source ID \\
    dr2\_source\_id & \ldots{} & int64 & \gaia{} DR2 source ID \\
    tic\_id & \ldots{} & int64 & \tess{} Input Catalog (TIC) ID (TIC~8.2) \\
    match\_dr3\_source\_id & \ldots{} & int64 & \gaia{} DR3 source ID of matched \blancoone{} member \\
    ra & deg & float64 & \gaia{} DR3 right ascension (J2016.0) \\
    ra\_error & deg & float64 & \gaia{} DR3 right ascension uncertainty (J2016.0) \\
    dec & deg & float64 & \gaia{} DR3 declination (J2016.0) \\
    dec\_error & deg & float64 & \gaia{} DR3 declination uncertainty (J2016.0) \\
    pmra & \si{\mas\per\year} & float64 & \gaia{} DR3 proper motion in right ascension (J2016.0) \\
    pmra\_error & \si{\mas\per\year} & float64 & \gaia{} DR3 proper motion uncertainty in right ascension (J2016.0) \\
    pmdec & \si{\mas\per\year} & float64 & \gaia{} DR3 proper motion in declination (J2016.0) \\
    pmdec\_error & \si{\mas\per\year} & float64 & \gaia{} DR3 proper motion uncertainty in declination (J2016.0) \\
    phot\_g\_mean\_mag & mag & float32 & \gaia{} DR3 \gaiaG{} magnitude \\
    bp\_rp & mag & float32 & \gaia{} DR3 \BPminusRP{} color \\
    phot\_g\_abs\_mag & mag & float32 & \gaia{} DR3 absolute \gaiaG{} magnitude, using \gaia{} DR3 parallax \\
    teff & K & float32 & Stellar effective temperature derived via the relation of \citet{2020ApJ...904..140C} (\autoref{sec:teff}) \\
    has\_cdips & \ldots & bool & Whether the target has a valid CDIPS light curve \\
    ls\_period & d & float64 & Rotation period determined by Lomb--Scargle periodogram \\
    ls\_flags & \ldots{} & string & Flags used in the vetting process (\autoref{sec:ls}) for Lomb--Scargle periods \\
    \enddata 
    \tablecomments{The full table is available online in machine-readable format.
    The column descriptions are included here to demonstrate the table's content.}
    \tablereferences{\gaia{} DR3: \citelindengrengaiadrthree.
    \gaia{} DR2: \citegaiatwo.
    TIC~8.2: \citet{tic8,tic82}.
    }
\end{deluxetable*}

\begin{figure*}
    \centering
    \includegraphics[width=\textwidth]{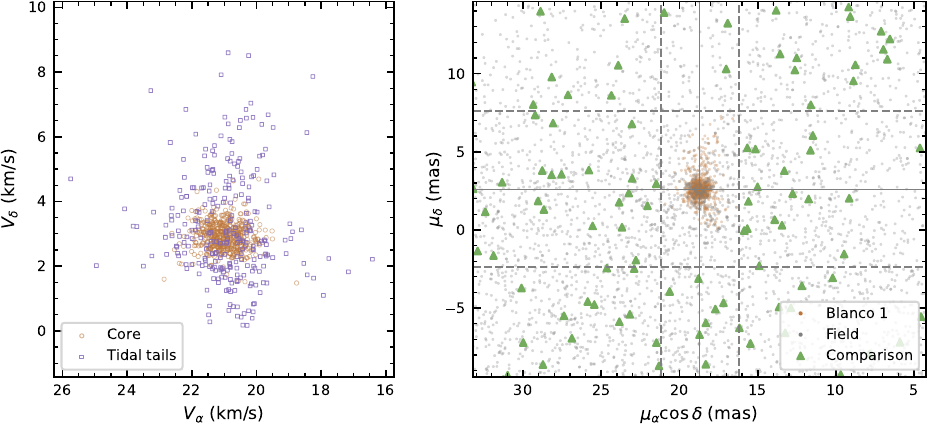}
    \caption{\textbf{Left:} Proper motion of Blanco 1 member candidates,
    converted into physical units using the corrected distances by \citet{2021ApJ...912..162P}. 
    The circular orange marks are the stars in the cluster core,
    and the square magenta marks are those in the tidal tails.
    \textbf{Right:}
    Exclusion in proper motion space for the comparison sample.
    The orange marks are \blancoone{} candidates,
    and the gray marks are field stars satisfying the first three criteria
    outlined in \autoref{sec:field_star} (i.e. before the exclusion in proper motion space).
    The triangular purple marks are stars in the selected comparison sample.
    A box of
        \SI{\pm 2.5}{\mas\per\year} in $\mu_\alpha$
        and \SI{\pm 5}{\mas\per\year} in $\mu_\delta$
    (shown in dashed gray lines)
    centered at $(\mu_\alpha, \mu_\delta) \approx (18.71, 2.61)\,\si{\mas}$
    (indicated by solid thin gray lines)
    is drawn to exclude potentially unidentified members of the \blancoone{} cluster from
    contaminating the sample.
    }
    \label{fig:proper_motion}
\end{figure*}

\subsection{Selecting a comparison sample of field stars}
\label{sec:field_star}

To assess whether the intrinsic rotation distribution of field stars might bias our interpretation of the candidate \blancoone{} members,
we constructed a comparison sample of field stars for experimental control.
Our selection procedure involved two steps.
In the first step, we aimed to construct a set of field stars
in the general vicinity of the \blancoone{} candidates.
We began by selecting stars
in the same \healpix{} \citep{2005ApJ...622..759G} level~4 pixels
from the \gaia{} DR3 catalog,
and then imposed
the following limits in photometric magnitude,
parallax, and proper motion:
\begin{enumerate}
    \item \gaia{} \gaiaG{} magnitude: 6.5--19.5.
    \item Parallax: \SIrange[range-phrase=--]{3.547}{5.205}{\mas}.
    \item Parallax relative error: $< 10$ per cent.
    \item Proper motion: \emph{exclude}
        the following region around median cluster proper motion
        (\pmra{}, \pmdec{}):
        $\approx (18.71 \pm 2.5, 2.61 \pm 5) \, \si{\mas\per\year}$.
\end{enumerate}
The exclusion in proper motion space is important, as it helps eliminate
unidentified members of \blancoone{} among the field stars
(\autoref{fig:proper_motion}).
Applying these criteria resulted in \num{23596} comparison sample candidates.

\begin{figure*}
    \centering
    \includegraphics[width=\textwidth]{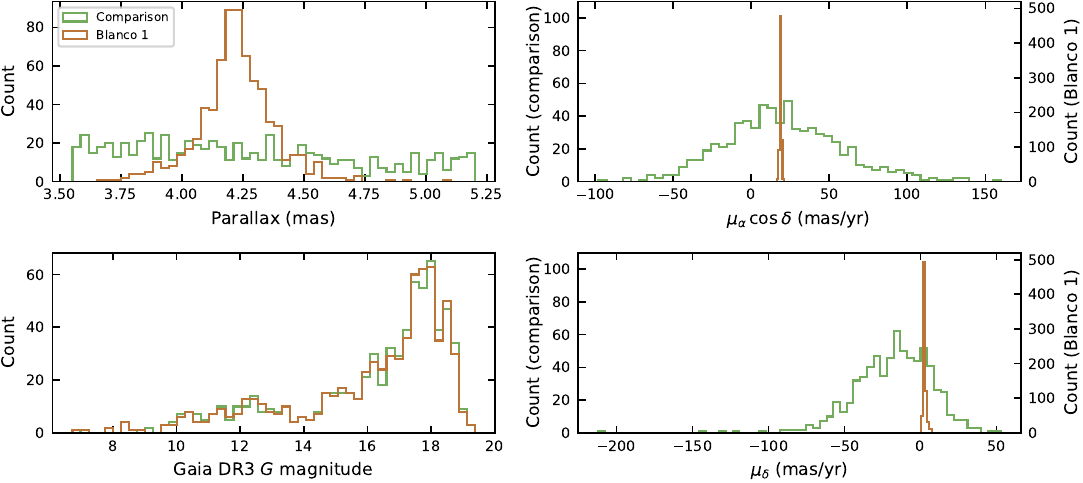}
    \caption{Histograms comparing the \blancoone{} member candidates (orange lines)
    and the selected comparison sample of field stars (purple lines).
    \textbf{Top left:} Parallax histograms. The \blancoone{} candidates shows a clustering in parallax,
    while the comparison sample is evenly distributed.
    \textbf{Top and bottom right:} Proper motion histograms. The \blancoone{} candidates show a sharp peak
    in proper motion space, and they are plotted on a different scale (right $y$-axis) for convenience. The comparison sample excludes those peaks; see also \autoref{fig:proper_motion}.
    \textbf{Bottom left:} Gaia DR3 \gaiaG{} magnitude histograms. The distributions for \blancoone{} candidates and the comparison sample closely match.
    }
    \label{fig:compare_control_hist}
\end{figure*}

In the second step,
we performed a one-to-one pairing between the \blancoone{} candidates 
and the comparison sample candidates.
For each \blancoone{} candidate,
we chose the most similar star within the same \healpix{} level~4 pixel
according to a similarity index that is the quadrature sum
of the difference in \gaia{} \gaiaG{}, \gaiaBP{}, and \gaiaRP{} magnitudes.
If a star had already been matched to a previous \blancoone{} candidate,
then we rejected it and moved on to the next most similar star.
This procedure resulted in a matched sample of field stars
that closely resemble \blancoone{} candidates in spatial, magnitude, and color distribution
(\autoref{fig:hr}, \autoref{fig:compare_control_hist}).
The selected field stars are listed in \autoref{tab:control}.

The effective temperatures of the comparison sample stars
are derived using the same procedure as for the \blancoone{} stars
in \autoref{sec:teff} and listed in \autoref{tab:control}.
While the comparison stars are expected to be older than the benchmark stars
used for the color--\tempeff{} relation,
our intention here is to calculate a temperature--rotation period relation
as if the comparison stars are field stars contaminating the \blancoone{} candidates.
Thus, even though the calculated \tempeff{} may not accurately reflect the true \tempeff{} of these likely older stars,
they are sufficient for the purpose of establishing a comparison sample.

\section{TESS Light Curves} \label{sec:cdips}

The procedures described above yielded lists of 703 candidate \blancoone{} members
and the same number of comparison field stars.
To measure the photometric rotation periods of these two stellar samples,
we used the full-frame images from the \tess{} prime mission \citep{2015JATIS...1a4003R}.
All of the stars were observed in either sectors~2 or 3 of the \tess{} prime mission,
which spanned \ut{} 2018 August 22 -- October 18.  
\tess{} observed full-frame images at a \SI{30}{\minute} cadence during its prime mission.

To make the light curves, we used the difference imaging pipeline
\citep{2019zndo...3370324B}
developed by the CDIPS project \citep{2019ApJS..245...13B}.
This pipeline performs forced-aperture photometry on difference images that are constructed on a sector-by-sector basis, by adding up the flux in circular apertures projected onto known stellar positions from Gaia DR2.
The reference stellar fluxes in the TESS band are computed based on \gaia{} DR2.
Since the default CDIPS target selection%
\footnote{See the MAST documentation
at \url{https://archive.stsci.edu/hlsp/cdips},
\unskip\dataset[10.5281/zenodo.10790463]{\doi{10.17909/t9-ayd0-k727}}.}
did not include either the \citet{2021ApJ...912..162P} or the comparison field stars,
we performed the photometry using difference images that were cached during the initial reductions of these fields.
This procedure yielded 603 light curves for the cluster sample
and 668 light curves for the comparison sample.%
\footnote{These light curves are available on Zenodo at \unskip
\dataset[10.5281/zenodo.10790463]{\doi{10.5281/zenodo.10790463}}.}
The remaining stars did not fall on the CCD sensors during the \tess{} prime mission.
After considering the median point-to-point scatter of the light curves of the cluster members as a function of \tess{} magnitude,
we adopted circular apertures with a radius of 1.5~pixels for stars brighter than 
a \emph{TESS} magnitude of $13.3$,
and the smaller 1-pixel-radius circular apertures for fainter stars.
This process yielded raw image-subtracted light curves (\texttt{IRM}),
light curves corrected with a principal component analysis cotrending approach
described in Appendix~B of \citet{2021AJ....162..197B}
(\texttt{PCA}), as well as measurements of any
residual flux in a local annulus around each circular aperture (\texttt{BGV}).
By default, we used the \texttt{PCA} time series when measuring
rotation periods, but we used all three datasets to visually assess their
validity, as described below.

\section{Estimating Rotation Periods} \label{sec:period}

\subsection{Lomb--Scargle periodograms}
\label{sec:ls}

\begin{table}
    \caption{Flags used in the vetting process for Lomb--Scargle periods.}
    \label{tab:vetting_flags}
    \centering
    \begin{tabular}{cl}
        \hline
        Flag & Description  \\
        \hline
        A & Period aliases present / ambiguous period \\
        B & Possible blend or eclipsing binary \\
        D & Detrending (PCA) artefacts \\
        F & Possible spikes or flares \\
        G & Background (BGV) correlation \\
        I & Affected by instrumental effects / scattered light \\
        M & Marginal amplitude \\
        V & Variable star / stellar activity \\
        \hline
    \end{tabular}
\end{table}

We adopted Lomb--Scargle \citep[LS;][]{1976Ap&SS..39..447L,1982ApJ...263..835S} periodograms
as implemented by \textsc{Astropy} \citep{astropy:2013,astropy:2018,astropy:2022}
as the main method for detecting rotation periods.
For each sector's light curve of \blancoone{} candidates and comparison samples,
we calculated LS periodograms
over a frequency grid ranging from
\SIrange[parse-numbers=false]{1/27.4}{24}{\per\day},
with the lower limit corresponding to one sector of \tess{} observations
and the upper limit the Nyquist frequency of the \SI{30}{\minute} cadence.
The grid spacing was chosen adaptively by \textsc{Astropy} to have
roughly 5 grid points per significant periodogram peak.
We also generated a LS periodogram for the window function,
which takes a value of unity for each observed sample.
For each sampled periodogram, we recorded the five highest local maxima
as the five most significant peaks,
and calculated their false-alarm probability with the bootstrap method
\citep{2014sdmm.book.....I,2018ApJS..236...16V}.

We then manually vetted the detected periodogram peaks in order to identify the most likely true period
and to eliminate insignificant peaks (false-alarm probability of $4\sigma$).
For reference during the vetting process,
we also generated phase-folded light curves (\texttt{PCA} time series) at the five most significant periodogram peaks,
as well as the unfolded \texttt{PCA}, \texttt{IRM}, and \texttt{BGV} time series,
in order to identify light curves suffering from instrumental or systematic anomalies
and exclude them from detected periods.
Among stars with a valid CDIPS light curve,
we detected likely rotation periods for 47\% of the candidate tidal tail members and 62\% of the core members,
while the comparison sample of field stars yielded rotation period detection for only 28\%.
As \autoref{fig:hr} shows,
we detected rotation periods in fewer proportion of stars
towards the cooler end of the HR diagram.
Since these stars are dimmer in apparent and absolute brightness,
it is likely that the amplitudes of their rotation signals, if present, fall below the detection threshold of \tess{}.

\paragraph{Flags used in manual vetting}
In the vetting process, we made use of several flags as summarized in \autoref{tab:vetting_flags}.
Detailed descriptions of the flags are as follows.
\begin{enumerate}
    \item \textbf{A} -- The periodogram contains multiple peaks that correspond to plausible phase-folded light curves. These peaks are often the true period and its first harmonic (period ratio 2:1), but can also be multiple peaks for stars showing complicated oscillations.
    \item \textbf{B} -- The sources either shows variations that resemble those of eclipsing binaries, or there are nearby sources in the target list that show variations at roughly the same period, indicating a possible blend.
    \item \textbf{D} -- The detrended time series (\texttt{PCA}) shows artifacts that are not present in the original (\texttt{IRM}) time series.
    \item \textbf{F} -- The light curve shows a flare-like systematic, where the magnitude increases sharply by more than 0.5 magnitudes and then drops down to the original magnitude within 1~day.
    \item \textbf{G} -- The original or detrended time series show correlation with the background flux values (\texttt{BGV} column in the FITS file). In this case, the first peak that does not correspond to variations in \texttt{BGV} (i.e. due to end-of-orbit scattered light) is selected. If all detected periodogram peaks can be entirely explained by this correlation, then the source is marked as no valid period detected.
    \item \textbf{M} -- The detected rotational signal is marginal. The amplitude is comparable to the scatter.
    \item \textbf{I} -- The light curve is affected by instrumental effects, such as scattered light.
    \item \textbf{V} -- The light curve shows stellar activity or variability that deviates from a regular sinusoid.
\end{enumerate}
These flags are reported in \autoref{tab:blanco1} and \autoref{tab:control}.
We found, however, that excluding targets from later analysis based on the flags here
do not significantly alter the resulting the temperature--rotation period distribution (\autoref{sec:color_period}),
so we report these flags here largely in the interest of transparency
and in the hope that it will be useful to future studies of \blancoone{}.

\subsection{Gaussian process regression}
\label{sec:gp}

We used a Gaussian process (GP) to model quasi-periodic covariance in the light curve time series.
In constructing the GP model and sampling the posterior distribution of its parameters,
we aim to estimate the uncertainty of the detected rotation period in the light curves.
Further motivation for adapting GP regression to measure stellar rotation periods is described 
in the previous rotation period study of \blancoone{}
by \citet{2020MNRAS.492.1008G},
which built on the work by \citet{2018MNRAS.474.2094A}.
Here, we broadly follow the method of \citet{2020MNRAS.492.1008G}
and adopted
\texttt{RotationTerm} 
from the Python package \celeritetwo{}
\citep{celerite1,celerite2}
as our GP kernel,
and we include its full definition in this section for the sake of completeness.
Unlike \citet{2020MNRAS.492.1008G}, however,
we omit their non-periodic term in the GP kernel
since our chosen light curve time series (\texttt{PCA}; see \autoref{sec:cdips})
has already been detrended.

The \texttt{RotationTerm} kernel is the mixture of two simple harmonic oscillator (SHO) terms,
each with a power spectrum density (PSD)
\begin{equation}
  \mathcal{S}[S_0, Q, \omega_0](\omega) = \sqrt{\frac{2}{\pi}}
    \frac{S_0 \, \omega_0^4}{\big( \omega^2 - \omega_0^2 \big)^2 + \omega_0^2 \omega^2 / Q^2},
\end{equation}
where $\omega$ is the angular frequency argument of the PSD
and amplitude $S_0$, quality factor $Q$, and resonant frequency $\omega_0$ are parameters of the GP term
(\citealt*{1990ApJ...364..699A}; \citealt{celerite1}).
The two SHO terms are at the fundamental period $P$ and first harmonic (half) period $P/2$.
For the ease of interpretation, these parameters are expressed in terms of
a standard deviation-like parameter $\sigma$,
base quality factor $Q_0$,
difference in quality factor between the fundamental and first harmonic $\delta Q$,
and fractional power of the first harmonic $f$,
such that for the PSD of the two SHO terms
$\mathcal{S}[S_1, Q_1, \omega_1]$ and
$\mathcal{S}[S_2, Q_2, \omega_2]$,
\begin{subequations}
\begin{align}
    A &= \frac{\sigma^2}{1 + f}, \\
    S_1 &= \frac{A}{\omega_1 Q_1},
    & S_2 &= \frac{f A}{\omega_2 Q_2} , \\
    Q_1 &= Q_2 + \delta Q ,
    & Q_2 &= \frac{1}{2} + Q_0, \\
    \omega_1 &= \frac{4\pi Q_1}{P \sqrt{4 Q_1^2 - 1}},
    & \omega_2 &= 2 \omega_1 .
\end{align}
\end{subequations}
This parameterization has the advantage that if
$Q_0 > 0$ and $\delta Q > 0$,
then the GP is guaranteed to be underdamped
and the fundamental term has higher quality than the first harmonic.
Because the GP is otherwise centered on $0$,
it is offset by a mean term $\mu$.

We used \pymc{} \citep{2015arXiv150708050S,pymc2023} to sample the posterior of the GP parameters.
\pymc{} uses the No-U-Turn Sampler \citep{JMLR:v15:hoffman14a}
to perform Hamiltonian Monte Carlo \citep{1987PhLB..195..216D}.
We set the prior distributions for the \texttt{RotationTerm} parameters as
\begin{subequations}
\begin{align}
    \ln(P/\text{d}) &\sim \normaldist(\ln(P_\text{LS}/\text{d}), 0.2), \\
    \ln \sigma &\sim \normaldist(\text{standard deviation of magnitude}, 2), \\
    \ln Q_0 &\sim  \uniformdist(-50, 50), \\
    \ln \delta Q &\sim  \uniformdist(-50, 50), \\
    f &\sim \uniformdist(0, 1), \\
    \mu &\sim \normaldist(\text{median magnitude}, 1),
\end{align}
\end{subequations}
where $\uniformdist(a, b)$ is the uniform distribution on the interval $(a, b)$
and $\normaldist(m, s)$ is the normal distribution with mean $m$ and standard deviation $s$.
The median posterior value and the $1\sigma$ uncertainty corresponding to
the 68\% highest density interval
of the GP period
is listed in \autoref{tab:blanco1} for \blancoone{} candidates.

\section{Verifying Blanco 1's tidal tails with gyrochronology} \label{sec:results}

\subsection{Temperature--rotation period distribution} \label{sec:color_period}

\begin{figure*}
    \centering
    \includegraphics[width=\textwidth]{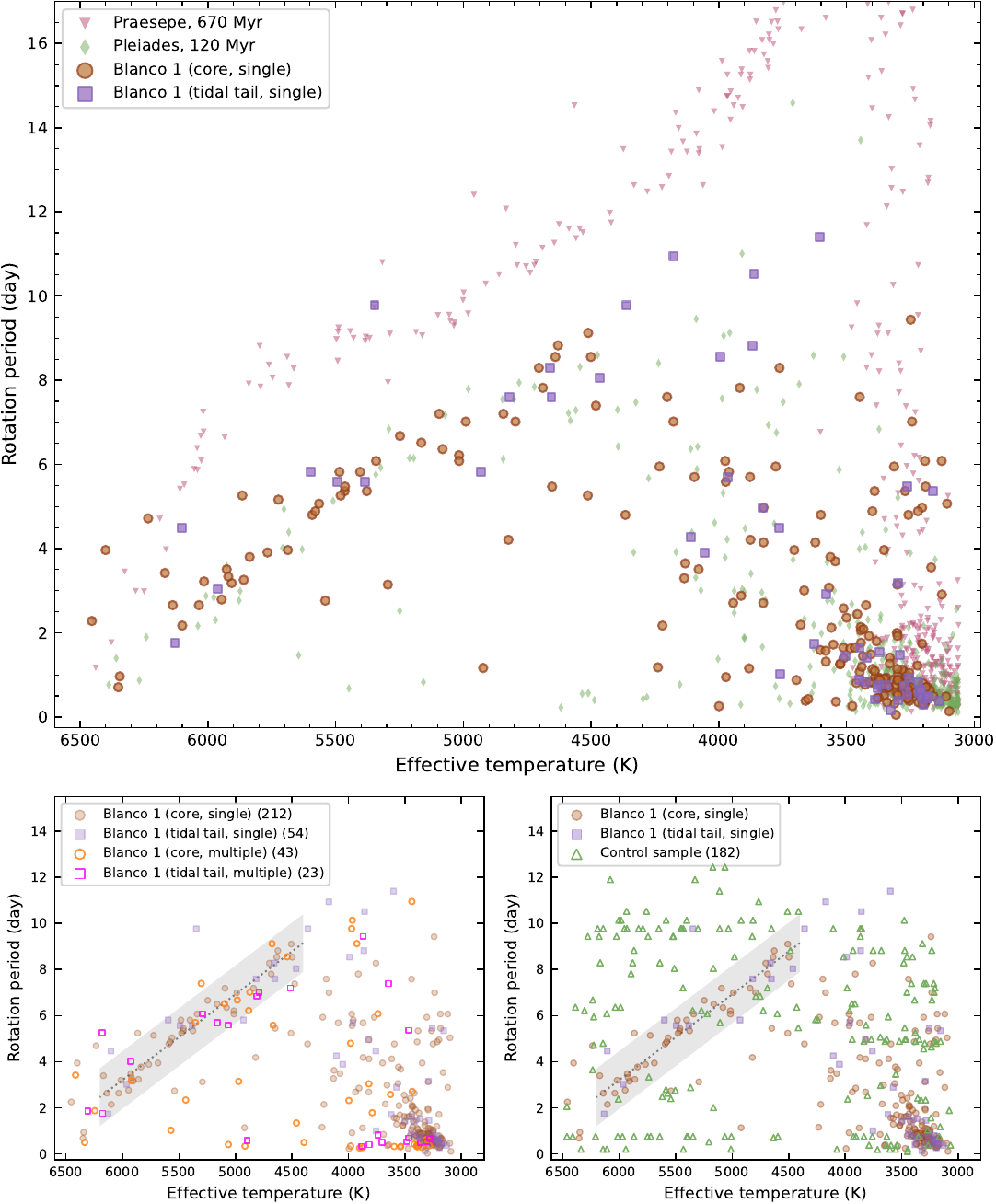}
    \caption{Temperature--rotation period diagrams.
    \textbf{Top:} Apparently single stars in \blancoone{} and benchmark comparison clusters.
    \textbf{Bottom left:} Single and multiple stars in \blancoone{}.
    \textbf{Bottom right:} \blancoone{} and comparison sample stars.
    For \blancoone{},
    the circular marks are core stars that fall within the tidal radius estimated by \citet{2021ApJ...912..162P},
    and the square marks are those that are outside the tidal radius, i.e. in the tidal tails.
    The effective temperatures are derived from \gaia{} DR2
    \BPminusRP{} colors using an empirical relation derived by \citet{2020ApJ...904..140C}.
    The gray dashed lines in the bottom panels
    indicate the gyrochrone of \blancoone{},
    with the shaded region being the $\pm \SI{1.25}{\day}$ band where 
    we consider a star to fall within the gyrochrone.
    }
    \label{fig:colorperiod}
\end{figure*}

The \blancoone{} candidates ranging from mid-F to mid-K dwarfs
($\SI{4400}{\kelvin} \leq \tempeff < \SI{6200}{\kelvin}$)
shows a well-defined temperature--rotation period, or gyrochrone, sequence.
\autoref{fig:colorperiod} compares the
temperature--period relation of single-star \blancoone{} candidates
in \autoref{fig:colorperiod},
to members of
the similarly aged Pleiades cluster
and the older Praesepe cluster.
In constructing \autoref{fig:colorperiod},
we estimated stellar effective temperatures from \gaia{} DR2 \BPminusRP{} color
using the method described in \autoref{sec:teff} for all three clusters,
and we use the rotation period measurements from
\citet{Rebull:2016} for the Pleiades
and \citet{2021ApJ...921..167R} for Praesepe, respectively.

As expected, \blancoone{}'s gyrochrone sequence,
which we visualize as the shaded band around the gray dashed line
in the bottom panels of \autoref{fig:colorperiod}),
broadly overlaps the Pleiades
and lies far below Praesepe,
consistent with previous findings by
\citep{2014ApJ...782...29C,2020MNRAS.492.1008G}.
The M dwarfs, on the other hand, show a large scatter in their rotation periods,
consistent with many of them not having converged onto the zero-age main sequence (ZAMS).
There is a single outlier, \gaia{} DR3 2333895128547040768, among the FGK stars
($\tempeff \approx \SI{5300}{\kelvin}$)
in the tidal tails that has a rotation period (\SI{9.77}{\day})
clearly too long to be part of the gyrochrone.
After cross-checking its light curves in later \tess{} sectors 29 and 69,
we determined that the detected period is likely double the true period
$\approx \SI{5}{\day}$.
This high degree of conformance to the gyrochrone indicates
that the tidal tail candidates from \citep{2021ApJ...912..162P}
likely have a low rate of contamination rate from unrelated field stars.
We formally compute the posterior distribution of this contamination rate in \autoref{sec:contam}.

\subsection{Contamination in the tidal tails}
\label{sec:contam}

We estimate the contamination rate of the tail stars by comparing
how many stars in the core and tail conform to the gyrochrone.
In determining the gyrochrone, we restrict to targets from
\SIrange{4400}{6200}{\kelvin},
which excludes M dwarfs and lower-mass K dwarfs as well as stars with temperatures above the Kraft break.
Then, we iteratively fit a line to the gyrochrone using \blancoone{} stars
by rejecting stars that fall more than \SI{1.25}{\day} from the fitted line at each step.
Once the line fitting converges,
we count the fraction of stars falling within \SI{1.25}{\day} of the gyrochrone
for the cores stars ($g_\text{core}$) and the comparison sample ($g_\text{comp}$).
Assuming that all core stars are true members
and that the comparison sample may have a certain fraction of \blancoone{} stars,
we can calculate a probability $p$ that a star in the tidal tail is within the gyrochrone range
\begin{equation}
    p = (1 - c) g_\mathrm{core} + c g_\mathrm{comp}
    = c (g_\mathrm{comp} - g_\mathrm{core}) + g_\mathrm{core}
\end{equation}
given a contamination ratio $c$ for the tidal tails.
The probability that $k$ out of $n$ tidal tail stars fall within the gyrochrone given this probability $p$ is given by the binomial distribution
\begin{equation}
    \Pr(k; n, p) = \binom{n}{k} p^k (1 - p)^{n - k} .
\end{equation}
Assuming a flat prior for the contamination ratio $c \in [0, 1]$,
we can marginalize this probability distribution over $c$
to obtain its posterior distribution
\begin{equation} \label{eqn:posterior}
    \Pr(c; n, k) = \mathcal{C}^{-1}
    p^k (1 - p)^{n - k} ,
\end{equation}
where $\mathcal{C}$ is a normalization factor given by the unsigned definite integral
\begin{equation}
    \mathcal{C} = \left| \int_0^1 p^k (1 - p)^{n - k} \, \mathrm{d}c \right|
    = \int_{g_\mathrm{comp}}^{g_\mathrm{core}}
        p^k (1 - p)^{n - k} \, \mathrm{d}p .
\end{equation}
Thus, we arrive at the posterior distribution
for the contamination of the tidal tails.

\begin{figure}
    \centering
    \includegraphics[width=\columnwidth]{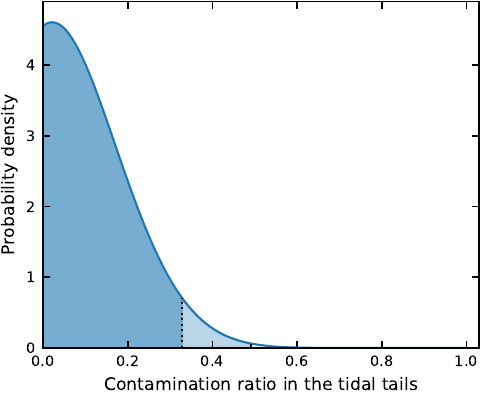}
    \caption{Posterior distribution for the contamination ratio of the tidal tail candidates in \blancoone{}.
    The shaded regions indicate the $2\sigma$ and $3\sigma$ intervals starting from zero,
    with the respective upper limits highlighted in black lines.}
    \label{fig:contam_pdf}
\end{figure}

Using the formalism established in the previous paragraph,
we measure gyrochrone fractions
$g_\text{core} = 0.741$ for the core stars and
$g_\text{comp} = 0.0847$ for the comparison sample.
There are $k = 16$ out of $n = 22$ tidal tail stars in the range of
$\SI{4400}{\kelvin} \leq \tempeff < \SI{6200}{\kelvin}$
that are in the gyrochrone.
Plugging these values into Equation \eqref{eqn:posterior}
yields a posterior distribution
\begin{equation} 
    \Pr(c) = \mathcal{C}^{-1}
    ( 0.741 - 0.656 c )^{16}
    ( 0.259 + 0.656 c)^{6}
\end{equation}
where $\mathcal{C}^{-1} \approx \num{1.828e6}$.
As visualized in \autoref{fig:contam_pdf},
the posterior is consistent with zero contamination,
and the $2\sigma$ and $3\sigma$ upper limits on the contamination ratio are
\contamratio{} and 0.49, respectively.

\begin{figure*}
    \centering
    \includegraphics[width=\textwidth]{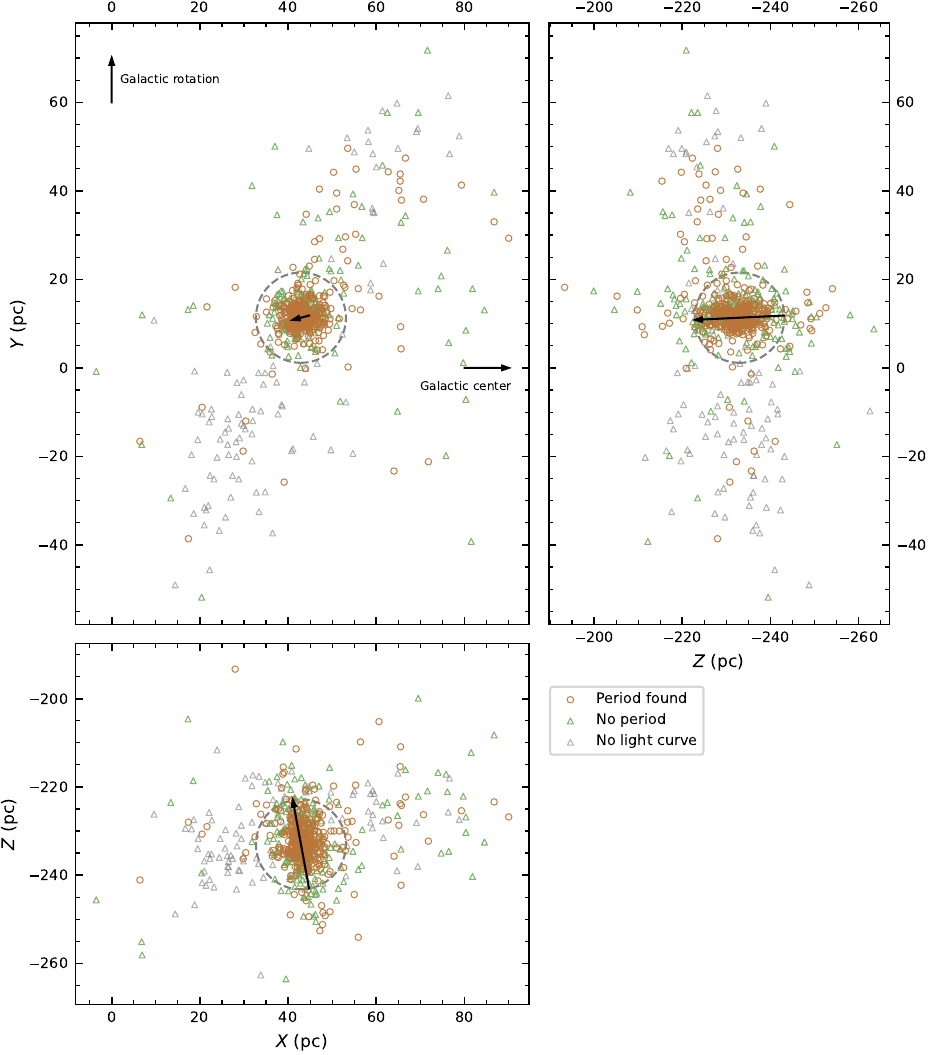}
    \caption{\blancoone{} members in Galactic Cartesian coordinates $XYZ$,
    as determined by \citet{2021ApJ...912..162P}.
    The dashed gray circle indicates the nominal tidal radius determined by \citeauthor{2021ApJ...912..162P},
    and the black arrow within the circle is the projected direction to the Sun.
    The circular orange marks are the stars with an actually measured rotation period,
    and the triangular gray marks are those without a measured period.
    }
    \label{fig:xyz}
\end{figure*}

\begin{figure*}
    \centering
    \includegraphics[width=\textwidth]{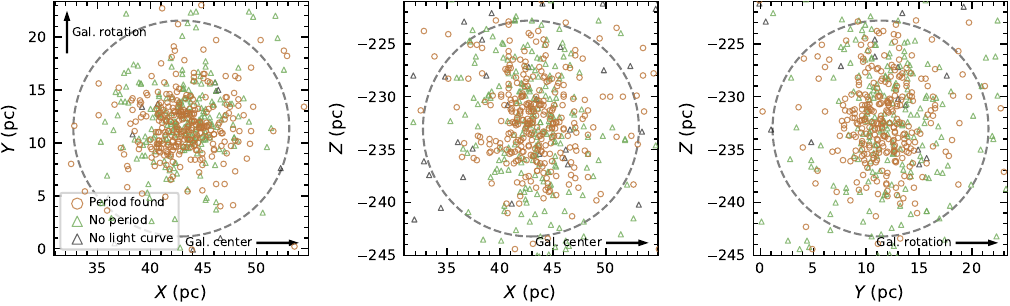}
    \caption{\blancoone{} core members in Galactic Cartesian coordinates $XYZ$,
    as determined by \citet{2021ApJ...912..162P}.
    The dashed gray circle indicates the nominal tidal radius determined by \citeauthor{2021ApJ...912..162P}
    The circular orange marks are the stars with a measured rotation period,
    and the triangular gray marks are those without a measured period.
    }
    \label{fig:xyz_core}
\end{figure*}

\section{Discussion} \label{sec:discussion}

\subsection{Confirmation of Blanco 1's tidal tails}

Our rotation period measurements serve to independently confirm
that the recently discovered tidal tails are truly associated with the
young open cluster \blancoone{}.
Previous work \citep{2021RNAAS...5..173L,2021A&A...645A..84M,2021ApJ...912..162P}
identified this extended structure
based on \gaia{} DR2 and DR3 spatio-kinematic data alone.
By checking that these new cluster member candidates conforms to the expected gyrochrone sequence of the core cluster members,
we have ruled out with high statistical confidence
that they can be explained by contamination from field stars (\autoref{sec:contam}).
In fact, the contamination is low enough
($< \contamratio{}$ at $2\sigma$)
to enable further studies, such as morphology, dynamics, and planetary occurrence rate, of \blancoone{}.

Now that we have confirmed the validity of its tidal tails,
we can revisit and confirm previous authors' conclusions
on the structure of \blancoone{} in the Galaxy.
Adopting the values of \citet{2021ApJ...912..162P},
\blancoone{} is located almost directly below the Galactic plane from the Sun (\autoref{fig:xyz}),
at a distance of approximately \SI{240}{\parsec}.
It has a tight core of $\approx 300$ stars, within a tidal radius of
\SI{10.2}{\parsec} (\autoref{fig:xyz_core}).
The spatial distribution of the core stars is elongated in the direction
almost perpendicular (inclined $\approx \ang{78}$) to the Galactic plane
(\autoref{fig:xyz}),%
\footnote{The line of sight to \blancoone{} is also almost perpendicular to the Galactic plane,
raising the question of whether this elongation is an artifact of inverting the parallax.
It turns out that \citet{2021ApJ...912..162P} already corrected for this effect through a Bayesian framework.
Tellingly, for the Coma Berenices cluster, which is located near the Galactic north pole,
\citeauthor{2021ApJ...912..162P} did not find such an elongation perpendicular to the Galactic plane,
which indicates that this elongation of the \blancoone{} core is likely real.
}
which \citet{2021ApJ...912..162P} hypothesized may be due to core stars
evaporating via the L1 and L2 Lagrange points in the Galactic tidal field,
similar to the $n$-body simulation results by \citet*{2008MNRAS.387.1248K}.

The reported extent of the cluster, however, is much larger.
Despite being unbound and much less dense,
the stars outside of the tidal radius contain $\approx 37\%$ of the mass of the cluster
\citep{2021RNAAS...5..173L}.
They protrude in opposite directions from the cluster core,
each extending \SIrange[range-phrase=--]{50}{60}{\parsec} from the cluster center.
The leading and trailing tails are slightly misaligned with the direction of
cluster rotation about the Galactic center,
consistent with them being the result of differential rotation
\citep{2021A&A...645A..84M}.
\subsection{How complete is the local young star census?}

Having confirmed \blancoone{}'s tidal tails 
using \tess{} photometric rotation periods,
we would like to know to what extent this strategy can be extended to other nearby young open clusters.
Answering this question would give us an idea of
how complete our current local young star census is
and thereby allow us to assess the future impact of these tidal tail stars
on the search for young exoplanets.
To this end,
we will make a back-of-the-envelope estimation of what fraction the presently identified tidal tail stars are
in terms of all nearby young stars where \tess{} is sensitive to detecting transiting planets.

We begin by considering how many stars have been identified in the literature
as candidates for the tidal tails of nearby young open clusters.
Querying the Montreal Open Clusters and Associations (MOCA) database
(Gagné et al., in preparation; \citealt{2024PASP..136f3001G,2018ApJ...856...23G}),
there are 32 open clusters younger than \SI{300}{\mega\year} within \SI{300}{\parsec}.
The MOCA database reports that 13 of these open clusters have detected tidal tails (\enquote{coronae}) in the literature.
Furthermore, the MOCA database reports that the core of these open clusters contain $\approx \num{10000}$ likely candidate members,
while their tidal tails contain $\approx 4600$ likely candidates.
Using \gaia{} $\gaiaRP < 13$ as a rough limit of \tess{} planet detection
narrows them down to $\approx 2800$ core versus $\approx 1100$ tidal tail candidates.

We then consider the number of all nearby young stars that are within \tess{}'s planet detection limit.
A quick query of the \gaia{} DR3 catalog yields
$\approx \num{7.2e6}$~sources with good parallax measurements
($\varpi / \sigma_\varpi > 10$)
within \SI{300}{\parsec},
$\approx \num{1.4e6}$ of which has $\gaiaRP{} < 13$.
Using a uniform age distribution of \SIrange[range-phrase=--]{0}{10}{\giga\year} as a zeroth-order approximation,
we arrive at $\approx \num{4.1e4}$~stars younger than \SI{300}{\mega\year} within this volume
whose transiting planets \tess{} is potentially sensitive to.
This implies that the presently known core and tidal tail candidates
respectively constitute $\approx 11\%$ and $\approx 3\%$ of stars
within \SI{300}{\parsec}
younger than \SI{300}{\mega\year}
whose potentially transiting planets \tess{} may be sensitive to.

\subsection{Searching for young exoplanets in tidal tails}

Even though open cluster candidates are a relatively small fraction of young nearby stars,
they have the advantage that their age is easily determined compared to field stars.
However, to date, we know relatively few planets in open clusters.
Predating \tess{} and \gaia{} DR2,
\citet{2018AJ....155..173C} listed $< 30$ planets discovered in clusters,
most of which were discovered by \kepler{} and \ktwo{}.
More recently,
\citet{2023AJ....166..219D} compiled a list of 73 confirmed planets and 84 planet candidates
cross-matched to the \circa $10^6$ stars in $\approx 8000$ groups identified by
\citet{2020AJ....160..279K} out to \SI{3}{\kilo\parsec}.
Considering that not all of these $\approx 8000$ groups are necessarily \enquote{open clusters} in the traditional sense,
this list of $\circa 100$ confirmed and candidate planets is a drop in the bucket
compared to the $\approx 4000$ likely candidate members of nearby young open clusters
amenable to transit detection by \tess{}.
Regardless of whether this dearth is the result of detection efficiency or intrinsic rarity,
expanding the number of target stars in well-dated open clusters will be crucial
to the search for more young planets.

In particular, tidal tail candidates can have an outsize impact in turning
previously assumed field stars with poor age constraints into open cluster members
with tight age constraints.
If we assume an average tidal dissolution timescale of $\approx \SI{100}{\mega\year}$
for open clusters,
then it becomes even more important to identify the dissolved cluster remnants with low spatial density
as cluster age increases.
As an extreme example, \cite{2021A&A...647A.137J} identified over 800 tidal tail candidates
of the Hyades open cluster spanning a spatial extent of \SI{800}{\parsec}.

The identification and confirmation of tidal tail candidates can thus have a real impact
on age determination for nearby young stars, and, by extension, young exoplanets.
As \citet{2021A&A...647A.137J} pointed out,
independent methods of verification such as rotation periods
are crucial in verifying candidates in a diffuse structure based on spatio-kinematic data alone,
since the membership probability from clustering algorithms is not sufficiently reliable.
If \blancoone{} can be used as a guide, we may double the number of targets for planet searches by including tidal tail candidates in addition to the core candidates.
Our calculation in \autoref{sec:contam} suggests that the contamination rate of these tidal tail candidates is probably low enough 
to enable occurrence rate studies of these young planets.

Thanks to its unique mission design,
\tess{} is well-positioned to contribute to the search for young planets in the tidal tails of
nearby young open clusters by both improving the local young star census and the transit observation coverage.
This paper demonstrates that \tess{} can improve the local young star census by verifying 
spatio-kinematic membership through stellar rotation period measurements.
Since young stars $< \SI{1}{\giga\year}$ are not expected to have rotation periods much longer than
one \tess{} spacecraft orbit of \SI{13.7}{\day},
the limited observational baseline of \tess{} sectors in its Prime Mission and first two Extended Missions
becomes irrelevant.
Thus, \tess{}'s all-sky coverage opens up all nearby open clusters for similar rotation period studies.
If we narrow our focus further to the relatively diffuse tidal tails,
we may entirely overstep the problem of crowding in \tess{}'s relatively big ($21''$) pixels for transit searches.
In summary, there remains deep potential for \tess{} to find yet more young planets in its continuing service.

\section{Conclusion}

In this study, we have used light curves derived from \tess{} full-frame images
to measure the rotation periods of member candidates in the young open cluster \blancoone{} and comparison field stars
and derived stellar effective temperature--rotation period diagrams for them.
Based on the gyrochrone sequence identified in the temperature--period diagrams,
we conclude that stars in the proposed tidal tails of \blancoone{} and
those in the core of the cluster are broadly consistent in age, with few outliers.
We detected rotation periods within \SI{1.25}{\day} of the gyrochrone
for 72\% of the candidate tidal tail members and 74\% of the core members
within a stellar effective temperature range of \SIrange{4400}{6200}{\kelvin},
while a comparison sample of field stars yielded
analogous rotation period detection for only 8.5\%.
Given these detection rates,
we find the contamination ratio for the tidal tails is consistent with zero
and $<\contamratio$ at the $2\sigma$ level,
confirming the existence of the tidal tails.
The relatively low contamination also suggests that finding more
tidal tails of known clusters and associations
is a viable strategy for
mapping out the dissolution of nearby open clusters
through a more complete local young star census.
Ultimately, this strategy could broaden
the search for new exoplanets in nearby young star clusters and associations,
leading to a determination of their occurrence rate.
\vskip\baselineskip
\noindent
The corresponding author would like to thank Michael V. Maseda,
in whose class this paper originated as a project.

This paper includes data collected by the \tess{} mission that are publicly available from the Mikulski Archive for Space Telescopes (MAST)
at\unskip\dataset[10.17909/t9-ayd0-k727]{https://dx.doi.org/10.17909/t9-ayd0-k727}.
We acknowledge the use of public \tess{} data from pipelines at the \tess{} Science Processing Operations Center (SPOC).
Resources supporting this work were provided by the NASA High-End Computing (HEC) Program through the NASA Advanced Supercomputing (NAS) Division at Ames Research Center for the production of the SPOC data products.
Funding for the \tess{} mission is provided by NASA's Science Mission Directorate.

This work has made use of data from the European Space Agency (ESA) mission
\href{https://www.cosmos.esa.int/gaia}{\gaia}, processed by the \gaia{}
Data Processing and Analysis Consortium
(\href{https://www.cosmos.esa.int/web/gaia/dpac/consortium}{DPAC}). Funding for the DPAC
has been provided by national institutions, in particular the institutions
participating in the \gaia{} Multilateral Agreement.

This research made use of the Montreal Open Clusters and Associations (MOCA) database, operated at the Montr\'eal Plan\'etarium (Gagn\'e et al., in preparation).

\vspace{5mm}
\facilities{Gaia, TESS}

\software{%
    Astropy \citep{astropy:2013, astropy:2018, astropy:2022},
    Astroquery \citep{2019AJ....157...98G},
    Celerite2 \citep{celerite1,celerite2},
    CDIPS pipeline \citep{2019zndo...3370324B},
    \href{https://git.ligo.org/lscsoft/ligo.skymap}{ligo.skymap},
    Matplotlib \citep{Hunter:2007,the_matplotlib_development_team_2023_10150955},
    Numpy \citep{harris2020array},
    Pandas \citep{mckinney-proc-scipy-2010,the_pandas_development_team_2023_10304236},
    PyMC,
    Scipy \citep{2020SciPy-NMeth}
}

\newpage

\bibliography{biblio}
\bibliographystyle{aasjournal}

\appendix

\section{Crossmatching candidates of Blanco~1 members}

{\catcode`\&=11
\gdef\citetalfonsoetal{\citet{2023A&A...677A.163A}}
\gdef\citettarricqetal{\citet{2022A&A...659A..59T}}
\gdef\citetmeingastetal{\citet{2021A&A...645A..84M}}
\gdef\citetcantatgaudinetal{\citet{2018A&A...618A..93C}}
\gdef\citetgaiacollabetal{\citet{2018A&A...616A..10G}}
}
\begin{deluxetable}{lcl}
    \savetablenum{A1}
    \tablecaption{Columns for the Table of \blancoone{} Candidates in Literature References.
    \label{tab:member_match}}
    \tablehead{%
        \multicolumn{1}{l}{Column} &
        \colhead{Data type} &
        \multicolumn{1}{l}{Description}
    }
    \startdata
    dr3\_source\_id & int64 & \gaia{} DR3 Source ID \\
    dr2\_source\_id & int64 & \gaia{} DR2 Source ID \\
    zhang\_num & int16 & Index column in Table~1 of \citet{2020ApJ...889...99Z} \\
    alfonso\_flag & bool & Flag for \citetalfonsoetal{} \\
    he\_flag & bool & Flag for \citet{2022ApJS..262....7H} \\
    tarricq\_flag & bool & Flag for \citettarricqetal{} \\
    pang\_flag & bool & Flag for \citet{2021ApJ...912..162P} \\
    meingast\_flag & bool & Flag for \citetmeingastetal{} \\
    zhang\_flag & bool & Flag for \citet{2020ApJ...889...99Z} \\
    cantat\_gaudin\_flag & bool & Flag for \citetcantatgaudinetal{} \\
    gaia\_collab\_flag & bool & Flag for \citetgaiacollabetal{} \\
    \enddata
    \tablecomments{Only the columns of this table are shown here to demonstrate its form and content.
    A machine-readable version of the full table is available online.
    The flag columns indicate whether a source is listed as a candidate member of \blancoone{}
    in the provided reference.}
\end{deluxetable}

In producing \autoref{fig:target_overlap},
we cross matched lists of candidate \blancoone{} members
by eight authors:
\citet{2018A&A...616A..10G},
\citet{2018A&A...618A..93C},
\citet{2020ApJ...889...99Z},
\citet{2021A&A...645A..84M},
\citet{2021ApJ...912..162P},
\citet{2022A&A...659A..59T},
\citet{2022ApJS..262....7H},
and \citet{2023A&A...677A.163A}.
The result of the cross-match is
\autoref{tab:member_match},
with distinct candidates indexed by their
\gaia{} DR3 and DR2 source IDs
and boolean flags indicating whether a candidate
is present a given reference.
As \citet{2020ApJ...889...99Z} did not include \gaia{} source IDs
in their table,
we included a column that gives their table's row index
if a candidate is present there.

In the final stages of preparing this paper,
we became aware of the work by
\citet{2023A&A...673A.114H},
which contains a membership list for \blancoone{}
as part of their comprehensive all-sky catalog of open clusters.
Although their list has more members ($N = 841$)
than \citet{2021ApJ...912..162P} ($N = 703$),
we still prefer the latter's for the analysis in this paper
as they provide corrected distances and the conversion to Galactic Cartesian coordinates.

\section{Main sequence of Blanco 1} \label{sec:ms_poly}

\begin{deluxetable}{cS}
    \savetablenum{B1}
    \tablecaption{Polynomial Coefficients for the Main Sequence of \blancoone{}.
    \label{tab:ms_poly}}
    \tablehead{%
        \colhead{Coefficient} &
        \colhead{Value}
    }
    \startdata
$c_0$ & 1.4827265 \\
$c_1$ & 7.283127 \\
$c_2$ & -33.80652 \\
$c_3$ & 99.9683 \\
$c_4$ & -138.4511 \\
$c_5$ & 106.2048 \\
$c_6$ & -48.087944 \\
$c_7$ & 12.809648 \\
$c_8$ & -1.8565603 \\
$c_9$ & 0.11289064 \\
    \enddata
\end{deluxetable}

\autoref{tab:ms_poly} gives the coefficients for an empirical polynomial fit
to the main sequence of \blancoone{}
in \gaia{} DR3 \BPminusRP{} colors
and \gaiaG{} magnitudes,
adopting the \blancoone{} candidates of \citet{2021ApJ...912..162P}.
These magnitudes are converted to the absolute scale
using the corrected distances by \citet{2021ApJ...912..162P}.
We iteratively fit the polynomial by excluding points more than 0.4 magnitudes away
from the predicted value at each step
until convergence.
The formula is in the form
\begin{equation}
    \hat{G}_\text{abs} = \sum_{k=0}^{9} c_k \, (\BPminusRP)^k ,
\end{equation}
where \BPminusRP{} is the \gaia{} DR3 color
and $\hat{G}_\text{abs}$ the predicted \gaia{} DR3 absolute magnitude.
The formula is valid in the range
$-0.1126 < \BPminusRP <  3.5895$,
but may not yield accurate results in the extreme blue end (due to a lack of samples)
or the red end (due to M dwarfs that have not converged to the zero-age main sequence).

Note that the coefficients are given up to 8 decimal places not as significant figures
but to reflect the machine precision of an IEEE 32-bit floating point number,
which is the data type of \gaia{} magnitudes.

\label{lastpage}

\end{document}